\documentclass[runningheads, citauthoryear]{llncs}
\usepackage{graphicx}
\usepackage{cite}
\usepackage{cleveref}
\usepackage{color}
\usepackage{xcolor}
\usepackage{listings}
\usepackage{tikz}
\usepackage{url}

\definecolor{lightgray}{rgb}{.9,.9,.9}
\definecolor{darkgray}{rgb}{.4,.4,.4}
\definecolor{purple}{rgb}{0.65, 0.12, 0.82}
\definecolor{darkgreen}{rgb}{0, 0.5, 0}
\definecolor{turquoise}{rgb}{0, 0.5, 0.5}
\definecolor{plum}{rgb}{.4, .14, .37}
\definecolor{mediumgreen}{HTML}{009900}
\definecolor{mediumred}{HTML}{CC0000}

\lstdefinelanguage{Solidity}{
  keywords={typeof, new, true, false, catch, function, struct, mapping, return, null, catch, switch, var, if, in, while, do, else, case, break},
  keywordstyle=\color{blue}\bfseries,
  ndkeywords={class, export, boolean, throw, implements, import, this, contract, library},
  ndkeywordstyle=\color{turquoise}\bfseries,
  keywords={[3]bool,address,uint, uint256, string},
  keywordstyle=[3]\color{darkgreen}\bfseries,
  identifierstyle=\color{black},
  sensitive=false,
  comment=[l]{//},
  morecomment=[s]{/*}{*/},
  commentstyle=\color{purple}\ttfamily,
  stringstyle=\color{red}\ttfamily,
  morestring=[b]',
  morestring=[b]"
}

\lstdefinelanguage{HoRSt}{
  sensitive = true,
  keywords = [1]{for, in, op, sel, expect, test, query, datatype, pred, clause, rule, let },
  keywords = [2]{match, with, store, select, mod },
  keywords = [6]{true, false, @T, @V, @D, @ADD, @STOP,@INVALID, @SELFDESTRUCT, 0, 1, 2 },
  keywords = [4]{int, bool, array, AbsDom, Opcode, CallData },
  keywords = [5]{MState, Exc, Halt, ReturnData },
  keywordstyle=[4]\color{plum}\bfseries,
  keywordstyle=[5]\color{darkgreen}\bfseries,
  keywordstyle=[6]\color{darkgreen!90!black}\bfseries,
  comment=[l]{//},
  morecomment=[s]{/*}{*/}
}

\newif\ifdraft
\draftfalse

\usepackage[users=CFM, , \ifdraft \else hide \fi]{uchanges}

\ChangesSetUserColor{C}{blue}
\ChangesSetUserColor{M}{red}
\ChangesSetUserColor{F}{darkgreen} 

\newcommand{\horst}{\emph{HoRSt}}
\newcommand{\ethor}{\emph{eThor}}
\newcommand{\java}{\emph{Java\textsuperscript{\texttrademark}}}
\newcommand{\smtlib}{\texttt{smt-lib}}
\newcommand{\zz}{\emph{z3}}
\newcommand{\souffle}{\emph{Souffl\'e}}
\newcommand{\fstar}{\emph{F}$^{*}$}
\newcommand{\javascript}{\emph{JavaScript}}
\newcommand{\solidity}{\emph{Solidity}}
\newcommand{\ethertrust}{\emph{EtherTrust}}
\newcommand{\hornc}{Horn}
\newcommand{\opcode}[1]{\textsf{#1}}
\newcommand*\circled[1]{\tikz[baseline=(char.base)]{
            \node[shape=circle,draw,inner sep=0.8pt] (char) {#1};}}
\newcommand{\rednumber}[1]{{\color{mediumred} \circled{#1}}}
\newcommand{\greennumber}[1]{{\color{mediumgreen} \circled{#1}}}

\begin{document}
\title{The Good, the Bad and the Ugly: \\ Pitfalls and Best Practices in Automated Sound Static Analysis of Ethereum Smart Contracts}
\titlerunning{The Good, the Bad and the Ugly}
%
\author{Clara Schneidewind\inst{1} \and
Markus Scherer\inst{1} \and
Matteo Maffei\inst{1}}
\authorrunning{Schneidewind et al.}
%
\institute{TU Wien, Vienna, Austria \\
\url{www.secpriv.wien} \\
\email{\{clara.schneidewind,markus.scherer,matteo.maffei\}@tuwien.ac.at}}
\maketitle              
\begin{abstract}
Ethereum smart contracts are distributed programs running on top of the Ethereum blockchain. Since program flaws can cause significant monetary losses and can hardly be fixed due to the immutable nature of the blockchain, there is a strong need of automated analysis tools  which provide formal security  guarantees. Designing such analyzers, however, proved to be challenging and error-prone. We review the existing approaches to automated, sound, static analysis of Ethereum smart contracts and highlight prevalent issues in the state of the art. Finally, we overview \ethor{}, a recent   static analysis tool that we developed following a principled design and implementation approach based on rigorous semantic foundations to overcome the problems of past works.
\end{abstract}
%
%
%
\section{Introduction}
 Blockchain technologies are revolutionizing the distributed system landscape, providing an innovative solution to the consensus problem leveraging probabilistic guarantees and incentives. In particular, they allow for the secure execution of payments, and more in general computations, among mutually distrustful parties. While some cryptocurrencies, like Bitcoin \cite{nakamoto2008bitcoin} provide only a limited scripting language tailored to payments, others, like Ethereum~\cite{wood2014ethereum}, support a quasi Turing complete\footnote{Supporting a Turing complete instruction set, Ethereum enforces termination by bounding the number of computation steps based on an prespecified resource limit.} smart contract language, allowing for advanced applications such as trading platforms~\cite{notheisen2017trading,mathieu2017blocktix}, elections~\cite{mccorry2017smart}, permission management~\cite{cruz2018rbac, azaria2016medrec}, data management systems~\cite{panescu2018smart, adhikari2017secure}, or auctions~\cite{hahn2017smart, galal2018verifiable}. 
With the growing complexity of smart contracts, however, also the attack surface grows. This is particularly problematic as smart contracts control real money flows and hence constitute an attractive target for attackers. In addition, due to the immutable nature of blockchains, smart contracts cannot be modified once they are uploaded to the blockchain, which makes the effects of security vulnerabilities permanent. This is not only a theoretical threat, but a practical problem, as demonstrated by infamous hacks, such as the DAO hack~\cite{thedao} or the Parity hacks~\cite{paritya,parityb} which caused losses of several millions of dollars.  
This state of affairs calls for reliable static analysis tools which are accessible to the users of the Ethereum system, that is, developers, who need to be able to verify their contracts before uploading them to the blockchain, and users interacting with existing smart contracts, who need tool assistance to assess whether or not those contracts (which are published in human unreadable bytecode format on the blockchain) are fraudulent. 

\subsubsection{State of the art.}
The bug-finding tool Oyente~\cite{luu2016making} (published in 2016) pioneered the (automatic) static analysis of Ethereum smart contracts. This work highlighted, for the first time, generic types of bugs that typically affect smart contracts, and proposed a tool based on symbolic execution  for the detection of contracts vulnerable to these bugs. 

A particular compelling feature of Oyente is that it  is a push-button tool that does not expect any interaction or deeper knowledge of the contract semantics  from the user. 
On the downside,  however, Oyente does not provide any guarantees on the reported results, being \NEWCfor{M}{200722}{neither sound (absence of false negatives) nor complete (absence of false positives)} \REMOVECfor{M}{200722}{affected by false positives and false negatives} and, thereby, yielding only a heuristic indication of contract security.
Given the importance of rigorous security guarantees in this context,  several approaches have been later proposed for the verification of Ethereum smart contracts. In particular, alongside tools that aim at machine-checked smart contract auditing~\cite{hildenbrandt_kevm:_2018, hirai2017defining, amani2018towards, bhargavan2016formal},  a line of work focused on  automation in the verification process to ensure  usability and broad adaption.
Despite  four years of intense research, however, until now only four works on such sound and fully automatic static analysis of Ethereum smart contracts have been published. Furthermore, all of these works exhibit shortcomings which ultimately undermine the security guarantees that they aim to provide. 

\subsubsection{Our contributions.}
Motivated by these struggles, we overview  the difficulties that arise in the design of sound static analysis for Ethereum smart contracts, and the pitfalls that so far hindered the development of such analyzers.
To this end, we first give a short introduction to the Ethereum platform and its native smart contract language (\Cref{sec:ethereum}). Afterwards, we illustrate the challenges in automated smart contract verification by overviewing existing works in this field with emphasis on their weak spots (\Cref{sec:related-work} and~\ref{sec:challenges}).
We conclude the paper summarizing our experiences in the design of \ethor{}~\cite{schneidewind2020ethor}, a static analyzer for EVM bytecode we recently introduced, along with a breakdown of the components we deem crucial to achieve efficiency and soundness (\Cref{sec:case-study}).

\section{Ethereum and the EVM Bytecode Language}
\label{sec:ethereum}
Ethereum is (after Bitcoin) the second most widely used cryptocurrency with a market capitalization of over 14 billion U.S.\ dollars\footnote{As of the fourth quarter of 2019, see \url{https://www.statista.com/statistics/807195/ethereum-market-capitalization-quarterly}.}. Compared to Bitcoin, Ethereum stands out due its expressive scripting language, that \REPLACECfor{FM}{200724}{allows for}{enables} the execution of arbitrary distributed programs, so called \emph{smart contracts}.
Ethereum smart contracts are stored on the blockchain in \REPLACEMfor{}{200531}{a} bytecode that is jointly executed by the network participants (also called \emph{nodes}) according to the Ethereum Virtual Machine (EVM) -- Ethereum's \REMOVECfor{FM}{200724}{smart contract} execution environment that is implemented in different clients\footnote{Currently, a Go, a C++ and a Python implementation are distributed by the Ethereum Foundation: https://github.com/ethereum/wiki/wiki/Clients,-tools,-dapp-browsers,-wallets-and-other-projects}.
Smart contract execution is part of Ethereum's consensus protocol:
The state of the system is determined by a jointly maintained, tamper-resistant public ledger, the \emph{blockchain}, that holds a sequence of transactions. Transactions do not only indicate money transfers, but can also trigger contract executions. To derive the \REMOVECfor{FM}{200724}{current }state of the system, every node locally executes the transactions in the blockchain as specified by the EVM.
For advancing the system, nodes broadcast transactions which are \REMOVECfor{FM}{200724}{then }assembled \REMOVECfor{FM}{200724}{by designated nodes (so called miners)} into blocks and appended to the blockchain. While the correctness of the system is ensured by the nodes validating all blocks, \REMOVECfor{FM}{200724}{the }fairness \REMOVECfor{FM}{200724}{of the system }is established by a proof of work mechanism: Only \REMOVECfor{FM}{200724}{such }nodes \NEWCfor{FM}{200724}{(called \emph{miners})} \REPLACECfor{FM}{200724}{being able to solve}{capable of solving} a computationally hard puzzle are eligible to propose a block. This \REPLACECfor{FM}{200724}{introduces}{adds} randomness \REPLACECfor{FM}{200724}{into}{to} the selection of proposers and \REMOVECfor{FM}{200724}{hence } prevents a minority from steering the evolution of the system.

\subsubsection{Ethereum Ecosystem} The state of the Ethereum system\REPLACECfor{FM}{200724}{, which we will refer to as \emph{global state},}{ (called \emph{global state})} consists of the state of all virtual accounts and their balances in the virtual currency \emph{Ether}. Smart contracts are special account entities (\emph{contract accounts}) that in addition to a balance hold persistent storage and the contract code. 
While non-contract (so called \emph{external}) accounts can actively transfer fractions of their balance to other accounts, contract accounts are purely governed by their code: A contract account can be activated by a transaction from another account and then \REPLACECfor{FM}{200724}{behaves exactly as defined by}{executes} its code, possibly transferring money or stimulating other contracts.  Similarly, a contract can be created on behalf of an external account or by another contract. We will in the following refer to such inter-contract interactions as \emph{internal transactions}, as opposed to \emph{external transactions} that originate from \REPLACECfor{FM}{200724}{an external account}{external accounts} and \REMOVECfor{FM}{200724}{that }are explicitly recorded on the blockchain.

\subsubsection{EVM Bytecode}
The EVM bytecode language supports designated instructions to account for the blockchain ecosystem. These encompass primitives for different flavors of money transfers, contract invocations, and contract creations. Most prominently, the \opcode{CALL} instruction allows for transferring money to another account while at the same time triggering code execution in case the recipient is a contract account.
Other domain-specific bytecodes include instructions for accessing the blockchain environment such as the instructions \opcode{SSTORE} and \opcode{SLOAD} for reading and writing the cells of the persistent contract storage. 
Further the instruction set contains opcodes for accessing information on the ongoing (internal) transaction {(its caller, input, or value transferred along)} and for computation: The EVM is a stack-based machine\REPLACECfor{FM}{200724}{, and hence supports}{ supporting} standard instructions for arithmetic and stack manipulation. The control flow of a contract is also subject to the stack-based architecture: conditional and unconditional jump instructions allow for resuming execution at another program position that is determined by the value \REMOVECfor{FM}{200724}{specified }on the stack.
While these features would make EVM bytecode Turing-complete, to enforce termination the execution of EVM smart contracts is bounded by an upfront-specified resource called \emph{gas}. Every instruction consumes \REMOVECfor{FM}{200724}{a certain amount of} gas \REMOVECfor{FM}{200724}{(potentially depending on the execution environment)} and the execution halts with an exception when running out of gas. The gas budget is set by the initiator of the transaction who will pay a compensation to the miner of the enclosing block for the effectively consumed amount of gas when executing the transaction.
Due to the low-level nature of EVM bytecode, smart contracts are generally written in high-level languages (most prominently the \solidity{}~\cite{solidity} language) and compiled to EVM bytecode.

\section{Related Work  on Automated  Sound Static Analysis of Ethereum Smart Contracts}
\label{sec:related-work}
We overview the state of the art in  the  automated static analysis of Ethereum smart contracts.
So far there have been works on four static analyzers published that come with (explicit or implicit) soundness claims: the dependency analysis tool Securify~\cite{tsankov_securify:_2018} for EVM bytecode, the static analyzer ZEUS~\cite{kalra_zeus:_2018} for \solidity{}, the syntax-guided \solidity{} analyzer NeuCheck~\cite{lu2019neucheck}, and the bytecode-based reachability analysis tool EtherTrust~\cite{GMS::CAV18}.
By implicit soundness claim,  we mean that the tool claims that a positive analysis result guarantees the contract's security (i.e., absence of false negatives with respect to a specific security property). 
\NEWCfor{M}{200722}{While Securify, ZEUS, and EtherTrust implement semantic-based analysis approaches, NeuCheck is purely syntax-driven.}

 Securify supports data and control flow analysis on the EVM bytecode level. To this end, it reconstructs the control-flow graph (CFG) from the contract bytecode and transforms it into SSA-form. Based on this structured format, it models immediate data and control flow dependencies using logical predicates and establishes datalog-style logical rules for deriving transitive dependencies which are automatically computed using the enhanced datalog engine \souffle{}~\cite{jordan2016souffle}.
For checking different security properties, Securify specifies patterns based on the  derived predicates which shall be sufficient for either proving a  property (compliance patterns) or for showing a property to be broken (violation patterns).

ZEUS analyzes \solidity{} contracts by first translating them into the intermediate language LLVM bitcode and then using off-the-shelf model checkers to verify different security properties.
In the course of the translation, ZEUS uses another intermediate layer for the \solidity{} language which introduces abstractions and that allows for the insertion of assumptions and assertions into the program code which express security requirements on the contract.
The security properties supported by ZEUS are translated to such assertion checks, possibly in conjunction with additional property-specific contract transformations.

NeuCheck analyzes \solidity{} contracts by pattern matching on the contract syntax graph. To this end, it translates \solidity{} source code into an XML parse tree. Security properties are expressed as patterns on this parse tree and are matched by custom algorithms traversing the tree.

 EtherTrust implements a reachability analysis on EVM bytecode by abstracting the bytecode execution semantics into (particular) logical implications, so called \hornc{} clauses, over logical predicates representing the execution state of a contract.
Security properties are expressed as reachability queries on logical predicates and solved by the SMT solver \zz{}~\cite{de2008z3}.
EtherTrust was a first prototype that later evolved into the  \ethor{} analyzer,  which we will discuss  in~\Cref{subsec:ethor}.

\NEWCfor{M}{200722}{All presented tools focus on generic (contract-independent) security properties for smart contracts. 
However, the underlying architectures allow for extending the frameworks with further properties. Foremost, the tool \ethor{} supports general reachability properties and hence also functional properties characterized in terms of pre- and postconditions. For the soundness considerations in this paper we put the focus on the abstractions of generic security properties.}

\section{Challenges in Sound Smart Contract Verification}
\label{sec:challenges}
 EVM bytecode
exposes several domain-specific subtleties that turn out to be challenging for static analysis.
\REPLACECfor{M}{200722}{
Furthermore, even the definition itself of security properties for smart contracts is highly non-trivial and subject to ongoing research.}
{Furthermore, characterizing relevant generic security properties for smart contracts is highly non-trivial and subject to ongoing research.} 
We will examine both of these problems in the following.

\subsection{Analysis Design}
 
We  summarize below the main challenges that arise when designing a performant and still sound analysis for Ethereum smart contracts:
\begin{itemize}
\item \emph{Dynamic jump destinations:}
Jump destinations are  statically unknown and computed during execution. They might be influenced by the blockchain environment as well as the contract state. As a consequence, the control flow graph of a contract is not necessarily determinable at analysis time.
\item \emph{Mismatch between memory and stack layout:}
The EVM has a (stack) word size of $256$ bits while the memory (heap) is fragmented into bytes and addressed accordingly. 
  Loading words from memory to the stack, and conversely writing stack values to memory, requires (potentially costly) conversions between these two value formats.

\item \emph{Exception propagation and global state revocation}
If an internal transaction (as, e.g., initiated by a \opcode{CALL}) fails, all effects of this transaction including those on the global state (e.g., writes to global storage) are reverted. However, such a failure is not propagated to the callee, who can continue execution in the original global state. Modeling calls must thus save the state before calling in order to account for global state revocation.
\item \emph{Native support for low-level cryptography:}
The EVM supports a designated \opcode{SHA3} instruction to compute the hash of some memory fraction. As a consequence, hashing finds broad adaption in Ethereum smart contracts, and  the Solidity compiler bases its storage layout on a hash-based allocation scheme. 
\item \emph{Dynamic calls:}
The recipient of an (inter-contract) call is specified on the stack and hence subject to prior computation. Consequently, the recipient {is not} necessarily {derivable} at analysis time, resulting in uncertainty about the behavior of the callee and the resulting effects on the environment.
\item \emph{Dynamic code creation:}
Ethereum supports the generation of new smart contracts during transaction execution: A smart contract can deploy another one at runtime. To do so, the creating smart contract reads the deployment code for the new contract from the heap. The newly created contract may hence be subject to prior computation and even to the state of the blockchain.
\end{itemize}


In order to effectively tackle these challenges, several  contributions of independent interest are required, such as domain-specific abstractions (e.g., suitable over-approximations of unknown environment behavior);
the preprocessing of the contract to reconstruct  its control flow or call graph;  (easily checkable) assumptions that restrict the analysis scope (e.g., restriction to some language fragment); 
and optimizations or simplifications in intermediate processing steps (e.g, contract transformations to intermediate representations).
Altogether, these analysis steps enlarge the semantic gap between the original contract semantics and the analysis,  making it harder to reliably ensure the soundness of the latter.
In the following, we will review in more detail the  tension between soundness and performance of the analysis, and how past works stumbled in this minefield.

\subsubsection{Soundness}
\label{subsubsec:correctness}
Ensuring the soundness of the analysis requires a rigorous specification of the semantics of EVM bytecode. 
The original semantics  was written  in the Yellow Paper~\cite{wood2014ethereum}. This paper however, from the beginning exhibited flaws~\cite{hirai2017defining, hildenbrandt_kevm:_2018, GMS::POST18} and underspecified several aspects of bytecode execution. The ultimate truth of smart contract semantics could therefore only be extracted from the client implementations provided by the Ethereum foundation. 
In the course of time, several formal specifications of the EVM semantics have been proposed by the scientific community~\cite{hildenbrandt_kevm:_2018, hirai2017defining,GMS::POST18}, leading the Yellow paper to be replaced by an executable semantics in the K framework~\cite{hildenbrandt_kevm:_2018}\footnote{Also called the Jello paper: \url{https://jellopaper.org}}.

For the high level language Solidity, despite first efforts within the scientific community~\cite{crafa2019solidity,zakrzewski2018towards,bartoletti2019minimal,yang2018lolisa,jiao2018executable}, 
there exists at present no full and generally accepted formal semantics. Consequently the semantics of Solidity is only defined by its compilation to EVM bytecode. Since the compiler is subject to constant changes, Solidity constitutes a moving target.

The complexity and uncertainty about the concrete semantics made most works build
 on ad-hoc simplified versions of the semantics which do not cover all language features and disregard essential aspects of the EVM's execution model. 

ZEUS~\cite{kalra_zeus:_2018}, for instance, defines an intermediate goto language for capturing the core of Solidity. The semantics of this language, however, is inspired by the (ad-hoc) semantic modeling used in Oyente~\cite{luu2016making}, inheriting an essential flaw concerning global state revocation: In case that an internal transaction returns \REPLACECfor{FM}{200724}{exceptionally}{with an exception}, the effects on the global state are not reverted as they would be in real EVM (and Solidity) executions.
Since the translation to the intermediate language is part of the analysis pipeline of~\cite{kalra_zeus:_2018}, such a semantic flaw compromises the soundness of the whole analysis.

Also Securify~\cite{tsankov_securify:_2018} introduces an ad-hoc formalism for EVM bytecode semantics. This {is not, however,} related to the dependency predicates used for the analysis, but just serves for expressing security properties. 
It is hence unclear to which extent the dependency predicates faithfully reflect the control flow and value dependencies induced by the EVM semantics. Assessing the correctness of this approach is difficult, since no full logical specification of the dependency analysis is provided\footnote{Only an excerpt is presented in~\cite{tsankov_securify:_2018}, and the public implementation at \url{https://github.com/eth-sri/securify} intermingles specification and implementation.}. Indeed we found empirical indication for the unsoundness of the dependency analysis in the presence of complicated control flow.
Consider the example contract depicted in~\Cref{fig:test-contract}.
\begin{figure}[t]
\begin{lstlisting}
contract Test {
  bool test = false;
  function flipper () { if (msg.sender != 0){flip();} }
  function flip () internal {test = !test;} }
\end{lstlisting}
\caption{Simple contract highlighting an unsoundness in Securify's dependency analysis.}
\label{fig:test-contract}
\end{figure}
For better readability, we present the contract in the high-level language \solidity{}, a language inspired by \javascript, that is centered around contracts which are used analogously to the concept of classes in object-oriented programming languages.
The depicted contract \lstinline|Test| has a global boolean field \lstinline|test|. Global fields are reflected in the persistent storage of the contract \REMOVECfor{FM}{200724}{account} and constitute the contract state. The public function \lstinline|flipper()| allows every account but the one with address $0$ to flip the value of the \lstinline|test| field: For checking the restriction on the calling account, the \lstinline|flipper()| function accesses the address of the caller using \solidity{}'s \lstinline|msg.sender| construct.
For writing the \lstinline|test| field, the internal function \lstinline|flip()| is called. Internal functions are not exposed to the public, but are only accessible by the contract itself and calls to such functions are compiled to local jumps. The use of internal functions consequently substantially complicates the control flow of a contract.

\REPLACECfor{M}{200722}{
We identified a soundness issue affecting the conditional reachability of contract locations.}{We identified a correctness issue that affects both the soundness and completeness of Securify.} This \REPLACECfor{M}{200722}{unsoundness}{incorrectness} becomes evident in the violation pattern that checks for unrestricted writes. An unrestricted write is a write access of the global storage that can be performed by any caller. The violation pattern states that such an unrestricted write is guaranteed to happen if there is a \opcode{SSTORE} instruction whose reachability does not depend on the caller of the contract. This pattern should not be matched by the \lstinline|Test| contract since the only write access to the \lstinline|Test|'s sole variable \lstinline|test| in function \lstinline|flip()| is only reachable via the function \lstinline|flipper| where it is conditioned on the caller (\lstinline|msg.sender|). Hence not every contract can write the \lstinline|test| variable, but the write access depends on the caller. Still Securify reports this contract to match the violation pattern, consequently proving the wrong statement that there is no dependency between writing the \lstinline|test| field and the account calling the contract.
Note that even though showing up in a violation pattern (hence technically producing false positives), the underlying issue \NEWCfor{M}{200722}{also} affects the soundness of the core dependency analysis\NEWCfor{M}{200722}{\footnote{We illustrate the issue with a violation pattern for easier presentation and since the affected compliance pattern turned out not to be implemented in Securify.}}.
\NEWCfor{M}{200722}{Securify specifies a \emph{may dependency} relation to capture (potential) abstract dependencies between program locations. For correctly (\REPLACEFfor{CM}{200724}{aka}{i.e.\ }soundly) abstracting the dependencies in real programs, the absence of a may dependency should imply a corresponding independence in the real program. Since the may dependency relation is used in both compliance and violation patterns, without such a guarantee Securify can be neither sound nor complete. The example refutes this guarantee and thereby illustrates the importance of providing clear formal correctness (\REPLACEFfor{CM}{200724}{aka}{i.e.\ }soundness) statements for the analysis.}

These two examples show how the missing semantic foundations of the presented analysis approaches can lead to soundness issues in the core analysis design itself.
These problems are further aggravated once additional stages are added to the analysis pipeline for increasing performance, since such additional stages are often not part of the correctness considerations.

\subsubsection{Performance}
\label{subsubsec:performance}
For performance reasons, it is often unavoidable to leverage well-established and optimized analysis frameworks or solvers. This leaves analysis designers with the challenge to transform their analysis problem into a format that is  expressible and efficiently solvable within the targeted framework while preserving the semantics of the original problem. 

 ZEUS~\cite{kalra_zeus:_2018} makes use of off-the-shelf model checkers for LLVM bitcode and hence requires a transformation of smart contracts  to LLVM bitcode. The authors describe this step to be a `faithful expression-to-expression translation' that is semantics preserving, but omit a proof for this property. The paper itself later contradicts this statement: The authors report on LLVM's optimizer impacting the desired semantics. This indicates that the semantics of the LLVM bitcode translation does not coincide with the one of the intermediate language, since it would otherwise not be influenced by (semantics-preserving) optimizations. 

The Securify tool~\cite{tsankov_securify:_2018} makes use of several preprocessing steps in order to make EVM bytecode amenable to dependency analysis: First it reconstructs the control flow graph of a contract and based on that transforms the contract to SSA form. The correctness of these  steps is never discussed. Indeed we found Securify's algorithm for control flow reconstruction to be unsound: The algorithm fails when encountering jump destinations that depend on blockchain information. In such a case the control flow should be considered to be non-reconstructable since a jump to such a destination may result in a valid jump at runtime or simply fail due to a non-existing jump destination. Securify's algorithm however does not report an error on such a contract, but returns a (modified) contract that does not contain jumps. Such an unsound preprocessing step again impacts the soundness of the whole analysis tool since it breaks the assumption that the contract semantics is preserved by preprocessing.

\subsection{Security Properties}
The Ethereum blockchain environment opens up several new attack vectors which are not present in standard (distributed) program execution environments. This is in particular due to  the contracts' interaction with the blockchain which is in general controlled by  unknown parties and hence needs to be \REMOVECfor{M}{200722}{often }considered hostile. It is a partly still open research question what characterizes a contract that is robust in such an environment.
A  well-studied property in this domain is  robustness against reentrancy attacks. We will focus on this property in the following to illustrate the challenges and pitfalls in proving a contract to be safe.

\subsubsection{Reentrancy Attacks}
\begin{figure}[t]
\includegraphics[width=\textwidth]{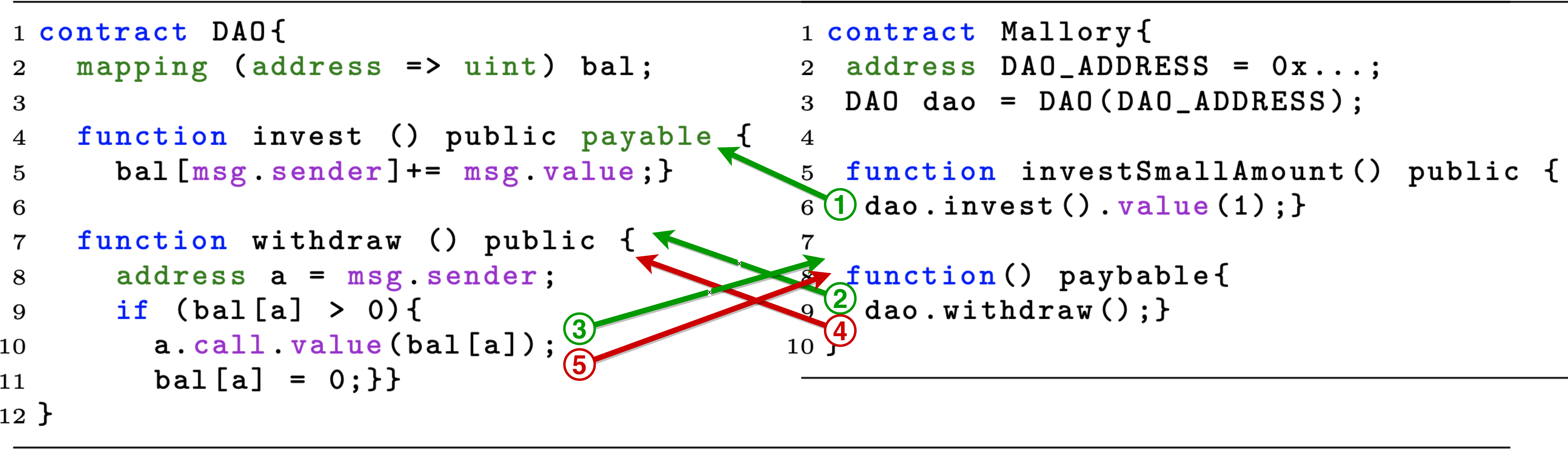}
\caption{Simplified DAO contract.}
\label{fig:dao}
\end{figure}

Reentrancy attacks became famous due to the DAO hack~\cite{thedao} in 2016 which caused a loss of over $60$ Million dollars, and ultimately led to a hard fork (a change in the consensus to explicitly ignore this particular incident) of the Ethereum blockchain.
The DAO was a contract implementing a crowd-funding mechanism which allowed users to invest and conditionally withdraw their invested money from the contract. An attacker managed to exploit a bug in the contract's withdraw functionality for draining the whole contract, stealing the money invested by other participants. 
We illustrate the main workings of this attack with a simplified example in~\Cref{fig:dao}.

The depicted \lstinline|DAO| contract has a global field \lstinline|bal| which is a mapping from account addresses to the payments that they made so far. The two (publicly accessible) functions of the contract allow arbitrary entities to invest and withdraw money from the contract. If an account with address $a$ calls the \lstinline|invest| function, the money transferred with this invocation is registered in the \lstinline|bal| mapping. 
Similar to \lstinline|msg.sender|, \solidity{} provides the variable \lstinline|msg.value| to access the value transferred with the currently executed (internal) transaction. The \lstinline|withdraw| function when being called by $a$, will check the amount of money invested by $a$ so far and in case of non-zero investments, transfer the whole amount of Ether (as recorded in \lstinline|bal[a]|) back to $a$. This is done using Solidity's \lstinline|call| construct for function invocations: it initiates a transaction to the specified address (here \lstinline|a|) and allows for the specification of the value to be sent along (using \lstinline|.value()|).
The attack on the \lstinline|DAO| contract can be conducted by an attacker that deploys a malicious contract \lstinline|Mallory| to first make a small investment to the \lstinline|DAO| contract (\greennumber{1}) that they later withdraw (\greennumber{2}). When the \lstinline|withdraw| function of the \lstinline|DAO| contract calls back to the sender (\lstinline|Mallory|, \greennumber{3}), not only the corresponding amount of Ether \REPLACECfor{FM}{200724}{will be}{is} transferred, but also code of \lstinline|Mallory| \REPLACECfor{FM}{200724}{will get}{is} executed. This is as in the case that not a specific (\solidity{}) function gets invoked with a contract call, the contract's \emph{fallback function} (a function without name and arguments) \REPLACECfor{FM}{200724}{will be}{is} executed. \lstinline|Mallory| implements this function \REPLACECfor{FM}{200724}{such that it calls}{to call} the \lstinline|DAO|'s \lstinline|withdraw| function (\rednumber{4}). Since at this point the balance of \lstinline|Mallory| in the \lstinline|bal| mapping has not been updated yet, another value transfer to \lstinline|Mallory| will be initiated (\rednumber{5}). By proceeding in this way, \lstinline|Mallory| can drain all funds of the \lstinline|DAO| contract.

The depicted attack is an example of how standard intuitions from (sequential) programming do not apply to smart contracts: In Ethereum one needs to consider that an internal transaction hands over the control to a (partly) unknown environment that can potentially schedule arbitrary contract invocations.

\subsubsection{Formalizing Security Properties}
\label{subsubsec:single-entrancy}
While bug-finding tools typically make use of heuristics to detect vulnerable contracts, there
have been two systematic studies that aim at giving a semantic characterization of what it means for a contract to be resistant against reentrancy attacks: The resulting security definitions are call integrity~\cite{GMS::POST18} and effective callback freedom~\cite{grossman2017online}. 

Call integrity follows non-interference-style integrity definitions from the security community. It states that two runs of a contract in which the codes of the environment accounts may differ, should result in the same sequences of observable events (in this case outgoing transactions). In simpler words, another contract should not be able to influence how a secure contract spends its money. Intuitively, this property is violated by the \lstinline|DAO| contract since an attacker contract can make the contract send out more money than in an honest invocation.

In contrast, effective callback freedom is inspired by the concept of linearizability from concurrency theory: It should be possible to mimic every (recursive) execution of a contract by a sequence of non-recursive executions\REMOVECfor{FM}{200724}{ of the very same contract}. The \lstinline|DAO| contract violates this property since the attack is only possible when making use of recursion (or callbacks respectively). After each callback-free execution, the \lstinline|investments| mapping will be updated, so that a subsequent execution will prevent further withdraws by the same party. 

While~\cite{GMS::POST18} shows how to over-approximate the hyperproperty call integrity by three simpler properties (the reachability property single-entrancy and two dependence properties),~\cite{grossman2017online} does not indicate a way of statically verifying effective callback freedom, but proves this property to be undecidable. This leaves sound, and (efficiently) verifiable approximations an open research question.

\subsubsection{Checking Security Properties}
\label{subsubsec:checking-security-properties}
The state-of-the-art sound analyzers discussed so far do not build on  prior efforts of semantically characterizing  robustness against reentrancy attacks, but come up instead with own ad-hoc definitions.

\paragraph{Securify}
 Securify expresses security properties of smart contracts in terms of compliance and violation patterns over data flow and control flow dependency predicates. In~\cite{tsankov_securify:_2018} it is stated that Securify supports the analysis of a property called `no writes after call' (NW) which is  different from (robustness against) reentrancy, but still aims at detecting bugs similar to the one in the DAO. 
The NW property is defined using an ad-hoc semantic formalism, and it states that for any contract execution trace, the contract storage shall not be subject to modifications after performing a \opcode{CALL} instruction.
\NEWCfor{FM}{200724}{Intuitively, this property should exclude reentrancy attacks by preventing that the guards of problematic money transfers are updated only after performing the money transferring call. However, this criterion is not sufficient e.g., since reentrancies can also be triggered by instructions other than \opcode{CALL}.} 
For proving \REPLACECfor{FM}{200724}{this}{the NW} property, the compliance pattern demands that a \opcode{CALL} instruction may not be followed by any \opcode{SSTORE} instruction. We found this pattern not to be sufficient for ensuring compliance  with the NW property (nor robustness against reentrancy).
We will illustrate this using a variation of the \lstinline|DAO| contract in~\Cref{fig:dao-lib}.
%
%
%
\begin{figure}[t]
\begin{lstlisting}[language=Solidity]
library Lib {
  struct Data { mapping (address => uint) map;}
  function write(Data storage self, address a, uint v) { self.map[a] = v;}
  function get(Data storage self, address a) returns (uint) {
    return (self.map[a]);} }

contract DAO {
  Lib.Data bal; 
  function invest() public payable { 
    Lib.write(bal, msg.sender, Lib.get(bal, msg.sender) + msg.value);}
  function withdraw () public {
    address a = msg.sender; 
    if (Lib.get(bal, a) > 0){
      a.call.value(Lib.get(bal, a));
      Lib.write(bal, a, 0);}} }
\end{lstlisting}
\caption{Simplified DAO contract using a library}
\label{fig:dao-lib}
\end{figure}
This contract implements the exact same functionality as the one in~\Cref{fig:dao}. The only difference is that the access to the balance mapping is handled via the library contract \lstinline|Lib|. Ethereum actively supports the use of library contracts in that it provides a specific call instruction, called \opcode{DELEGATECALL}, that executes another contract's code in the environment of the caller. When calling \lstinline|Lib.write| in the \lstinline|withdraw| function, such a delegated call to the (external) library contract is executed. Executing \lstinline|write| in the context of contract \lstinline|DAO| will then modify \lstinline|DAO|'s storage (instead of the one of the \lstinline|Lib| contract). In order to let the \lstinline|write| and the \lstinline|get| functionality access the right storage position (where \lstinline|DAO| stores the \lstinline|bal| mapping), these functions take as first argument the reference to the corresponding storage location.
Same as the version in~\Cref{fig:dao}, this contract is vulnerable to a reentrancy bug. Also, it violates the NW property: The storage of the contract can be changed after executing the call (when writing the \lstinline|bal|) mapping. 
Still, this contract matches the compliance pattern (which should according to~\cite{tsankov_securify:_2018} guarantee the contract to satisfy the NW property), since it does not contain any explicit \opcode{SSTORE} instruction.
This example illustrates how without a proven connection between a property and its approximation, the soundness of an analyzer can be undermined. This issue does not only constitute a singular case, but is a structural problem: There are counter examples for the soundness of $13$ out of the $17$ patterns presented in~\cite{tsankov_securify:_2018}, as we detail out in~\cite{schneidewind2020ethor}.

\paragraph{ZEUS}
 In~\cite{kalra_zeus:_2018}, the  property to rule out reentrancy attacks is only specified in prose as a function being vulnerable `if it can be interrupted while in the midst of its execution, and safely
re-invoked even before its previous invocations complete execution.'
This definition works on the level of functions, a concept which is only present on the \solidity{} level, and leaves open the question what it means for \emph{a contract} to be robust against reentrancy attacks. 
The authors distinguish between `same-function-reentrancy' and `cross-function-reentrancy' attacks, but do not consider cross-function reentrancy (where a function reenters another function of the same contract) in the analyzer. We found that without excluding cross-function reentrancy also single-function reentrancy cannot be prevented. 

Consider the  versions of the DAO contract depicted in~\Cref{fig:cross-functon-reentrancy} that aim to prevent reentrancy using a locking mechanism.
\begin{figure}[t]
\begin{minipage}[t]{0.49\textwidth}
%
%
\begin{lstlisting}[language=Solidity]
contract DAO{
  mapping (address => uint) bal; 
  uint lock = 0;
  function withdraw () public {
    if(lock ==1){throw;} @\label{line:lock-check1}@
    lock=1;
    address a = msg.sender; 
    a.call.value(bal[a]);
    bal[a] = 0;
    lock=0;}
  function switchLock () { 
    lock = 1-lock;} }
\end{lstlisting}
\end{minipage}%
\begin{minipage}[t]{0.49\textwidth}
%
\begin{lstlisting}[language=Solidity]
contract DAO{
  mapping (address => uint) bal; 
  uint lock = 0;
  function withdraw () public {
    if(lock ==1){throw;} @\label{line:lock-check2}@
    lock=1;
    address a = msg.sender; 
    a.call.value(bal[a]);
    bal[a] = 0;
    lock=0;} 
\end{lstlisting}
\end{minipage}
\caption{Simple versions of the DAO contract with reentrancy protection.}
\label{fig:cross-functon-reentrancy}
\end{figure}
The global \lstinline|lock| field tracks whether the \lstinline|withdraw| function was already entered (indicated by value \lstinline|1|). In that case, the execution of \lstinline|withdraw| throws an exception. Otherwise the \lstinline|lock| is set and only released when concluding the execution of \lstinline|withdraw|.
While the two depicted contracts implement the exact same \lstinline|withdraw| function, the first contract's function is vulnerable to a reentrancy attack, while the second one is safe. This is as the first contract implements a public \lstinline|switchLock()| function that can be used by anyone to change the \lstinline|lock| value. An attacker could hence mount the standard attack with the only difference that they would need to invoke the \lstinline|switchLock()| function once before reentering to disable the reentrancy protection in line~\ref{line:lock-check1}. Without exposing such functionality, the second contract is safe, since every reentering execution will be stopped in line~\ref{line:lock-check2}.
This example shows that ZEUS' approach of analyzing functions in isolation to exclude `same-function-reentrancy' is not sound.

Another issue in the reentrancy checking of ZEUS is caused by the reentrancy property exceeding the scope of the analysis framework. For proving a function resistant against reentrancy attacks, ZEUS checks whether it is ever possible to reach a call when a function is recursively invoked by itself. However, the presented translation to LLVM bitcode only models non-recursive executions of a function. Consequently, the reentrancy property cannot be expressed as a policy (which could be translated to assertions in the program code), but requires to rewrite the contract under analysis to contain duplicate functions that mimic reentering function invocations. 
This contract transformation is not part of any soundness considerations. As a result, not only the previously discussed unsoundness due to the lacking treatment of cross-function reentrancies is missed, but it is also disregarded that \solidity{}'s \lstinline|call| construct is not the only way to reinvoke a function. Indeed there are several other mechanisms (e.g., direct function calls) that allow for the same functionality. Still, ZEUS classifies contracts that do not contain an explicit invocation of the \lstinline|call| construct to be safe by default.

\paragraph{NeuCheck}
The NeuCheck tool formulates a syntactic pattern for detecting robustness against reentrancy attacks. The pattern checks for all occurrences of the \lstinline|call| function whether they are followed by the assignment of a state variable. As discussed for Securify, the absence of explicit writes to the storage does not imply that the storage stays unchanged. Hence the example in~\Cref{fig:dao-lib} would also serve as a counter example for the soundness claim of NeuCheck. Also, as discussed for ZEUS, \lstinline|call| is not the only way of invoking another contract, what reveals another source of unsoundness in this definition.
Furthermore, neither the security properties that the tool aims for are specified nor any justifications for the soundness of this syntactic analysis approach are provided.



\section{How to Implement a Practical, Sound Static Analysis?}
\label{sec:case-study}
After exposing the problems that can arise when designing a practical, sound static analysis, we discuss how we tackled them in  developing \ethor{} and the underlying  static analysis specification framework \horst{}~\cite{schneidewind2020ethor}. 
We then present the elements we identified as essential for designing an automated sound analysis: A semantic foundation, sound abstractions, and a principled implementation.

\subsection{Overview of \ethor{}}
\label{subsec:ethor}

\REMOVECfor{FM}{200724}{The design of }\ethor{} \REMOVECfor{FM}{200724}{and \horst{}} was preceded by an earlier prototype, called \ethertrust{}~\cite{GMS::CAV18}.
\ethertrust{} implemented the rules of a \REPLACECfor{FM}{200724}{formally specified}{formal} abstract semantics in \java{} and exported them to \zz{}.
While this design showed promising preliminary results, it turned out to be too inflexible for our purposes:
Changes in the abstract semantics had to be tediously translated to \java{} code\REMOVECfor{FM}{200724}{ which made it difficult to experiment with different approaches};
the non-declarative manner of specifying rules made them hard to write and review; and
the lack of a proper intermediate representation made it difficult to implement custom optimizations before passing the verification task to \zz{}.

These limitations are addressed by \horst{}~\cite{schneidewind2020ethor}, a dedicated high-level language for designing Horn clause based static analyses. \horst{} allows for the specification of Horn Clause based semantic rules in an \REMOVECfor{FM}{200724}{abstract,} declarative language and can apply different optimizations before translating them to \zz{} formulae.
Thus, the semantics specification and the tool implementation are logically separated and systematic experiments with different versions of the semantics are possible.
Additionally,  optimizations can be implemented independently from specific semantics, improving the overall performance in a \REMOVECfor{FM}{200724}{more }robust fashion.

\ethor{}~\cite{schneidewind2020ethor} combines \horst{} with a\REPLACECfor{FM}{200724}{ family of }{n }abstract EVM semantics, a parser to read EVM bytecodes, and a set of EVM-specific preprocessing steps, including the reconstruction of the control flow and the propagation of constants.  It supports general reachability analysis and in particular allows for \NEWCfor{FM}{200724}{(soundly)} verifying that a contract is single-entrant (following the definition in~\cite{GMS::POST18}).

\NEWCfor{FM}{200724}{What distinguishes \ethor{} from prior work discussed in~\Cref{sec:related-work} is its well defined analysis specification that is supported by rigorous formal soundness proofs, as well as its principle implementation design. Prior works do not come with thorough formalization and proofs what ultimately leads to soundness issues in the analyzers, as we confirmed empirically. In contrast, \ethor{} lives up to its theoretical soundness guarantees in an extensive evaluation while still being practical in terms of runtime and precision. In the following, we will discuss in detail the semantic foundations, modular design and implementation of \ethor{} as well as its empirical performance evaluation.}

\subsection{Semantic Foundations}

A formal soundness \REPLACECfor{FM}{200724}{statement}{guarantee} requires a formal semantics of the system under analysis.
\REPLACECfor{FM}{200724}{The specification of a semantics may happen in prose}{Such a semantics might be specified on paper}~\cite{wood_ethereum:2014} or in an executable fashion~\cite{hildenbrandt_kevm:_2018, GMS::POST18}, but in any case has to be precise enough to unambiguously capture all relevant aspects of the system.
While semantics defined in prose tend to be more readable, executable semantics lend themselves to automated \REPLACECfor{}{200529}{processing, such as checking for edge cases automatically\footnote{As mentioned in \cref{subsubsec:correctness}, even carefully reviewed semantics like \cite{wood_ethereum:2014} can leave certain behaviors underspecified.}, automated proofs of soundness and automated generation of further analysis tools, such as interpreters \cite{hildenbrandt_kevm:_2018}.}
{testing or tooling (e.g., the generation of interpreters or symbolic debuggers~\cite{hildenbrandt_kevm:_2018}).}
\ethor{} builds on the semantics presented in~\cite{GMS::POST18} which consists of a logical specification  as well as an executable \fstar{} semantics \NEWCfor{FM}{200724}{that was rigorously tested for its compliance with the Ethereum client software}.

Using a formal semantics,  security properties can be precisely characterized. \ethor{} bases its analysis for the absence of reentrancy attacks on the notion of single-entrancy~\cite{GMS::POST18}. Single-entrancy captures that the reentering execution of a contract should not initiate any further internal transactions. This property rules out reentrancy attacks and also contributes to the proof strategy for the more general call integrity property as detailed out in~\cite{GMS::POST18}.

\NEWCfor{FM}{200724}{
However, an executable semantics combined with precisely defined security properties alone does not yield a useful analysis tool.
While these components allow experts to semi-automatically verify contracts (using frameworks such as~\cite{hirai2017defining, GMS::POST18, hildenbrandt_kevm:_2018,amani2018towards}), automation generally requires abstractions to be feasible.}

\subsection{Sound Abstractions}
\REMOVECfor{FM}{200724}{
\REPLACECfor{FM}{200724}{Unfortunately, a}{A}n executable semantics combined with precisely defined security properties alone does not yield a useful analysis tool.
While these components allow experts to semi-automatically verify contracts (using frameworks such as~\cite{hirai2017defining, GMS::POST18, hildenbrandt_kevm:_2018,amani2018towards}), automation generally requires abstractions to be feasible.}

\REPLACECfor{FM}{200724}{When performance concerns or restrictions in the used analysis technique do not allow for an exact treatment, the target property needs to be over-approximated.}{
A first step to reduce the complexity of the analysis problem, and hence to make it amenable to automation, is to over-approximate the target property.}
For \ethor{}, we over-approximate the single-entrancy property by the simpler \emph{call unreachability}~\cite{GMS::CAV18} property.
Call unreachability breaks down single-entrancy to a simple criterion on the execution states of a single contract, as opposed to reasoning about the structure and evolution of whole call stacks.
Such over-approximations have to be proven sound \REMOVECfor{FM}{200724}{against the precise semantics} -- every program fulfilling the over-approximated property also has to fulfill the original property.
A corresponding proof for single-entrancy is conducted in~\cite{GMS::CAV18}.

\REPLACECfor{FM}{200724}{For efficiently verifying an (over-approximated) property}{To further simplify the analysis task}, the relevant parts of a contract's execution behavior need to be abstracted in a sound manner. \REPLACECfor{FM}{200724}{In the case of \ethor{} this is done by \REPLACECfor{FM}{200724}{formulating}{devising}}{In \ethor{} for this purpose we devised} an abstract semantics based on Horn clauses that \REPLACECfor{FM}{200724}{is proven}{we proved} in~\cite{schneidewind2020ethor} to soundly over-approximate the small step semantics in~\cite{GMS::POST18}. 
\REPLACECfor{FM}{200724}{In the following we will describe some of the abstractions that constitute the abstract semantics used in \ethor{}.}{The abstract semantics simplifies and summarizes complex execution scenarios that may emerge due to the uncertain blockchain environment, as we exemplify in the following.}

\REPLACECfor{FM}{200724}{
In the case of arbitrary execution environments (such as the blockchain)  the number of unknown environment inputs can make it}{
In the largely unknown blockchain environment it is} infeasible to track constraints on all unknown values.
Instead, following a standard technique in abstract interpretation, we enriched our domain of concrete computation values with a new value $\top$, signifying \emph{all} possible values. This designated symbol \REMOVECfor{FM}{200724}{is used to }over-approximate\NEWCfor{FM}{200724}{s} under-specified values while dropping \REMOVECfor{FM}{200724}{any }constraints on them.
Some computations, such as the SHA-3-computations or unaligned memory accesses in EVM, are \REPLACECfor{FM}{200724}{infeasible to model precisely and therefore are}{due to their complexity} over-approximated by $\top$ in \ethor{}.
\REMOVECfor{FM}{200724}{
Depending on the required level of precision and on the analyzed property, an abstract domain could of course be more fine-grained, e.g. incorporating a symbolic treatment of hash values.}

\REPLACECfor{FM}{200724}{Other aspects of the system may be plainly irrelevant and should be therefore dropped in order to simplify the static analysis.}{
Further, we abstract the initial invocation of a contract and its reentering executions as they might be scheduled by (unknown) contracts which are called during execution.}
In \ethor{} we ignore the call stack up until the first execution of the analyzed contract, and assume a contract to be called in an arbitrary environment instead.
Also\NEWFfor{M}{200722}{,} we only distinguish between the first execution of a contract under analysis in a call chain and a reentering execution of such a contract\REPLACECfor{FM}{200724}{, thereby collapsing}{. In this way we collapse} all reentering executions \NEWCfor{FM}{200724}{while modeling relevant storage invariants with a sophisticated domain-specific abstraction}.

\REMOVECfor{FM}{200724}{
Another aspect of the precise semantics that we ignore is gas consumption: as we have to assume (nearly) arbitrary gas budgets, ignoring them makes the analysis vastly simpler for only marginally reduced precision.}

For an extended discussion of the abstractions used in \ethor{}, including those for \REPLACECfor{FM}{200724}{calls between contracts}{inter-contract calls}\NEWCfor{FM}{200724}{, gas treatment,} and memory layout, we refer \REMOVECfor{FM}{200724}{the reader} to \cite{schneidewind2020ethor}.

\NEWCfor{FM}{200724}{In summary, \ethor{} provides a reliable soundness guarantee for the single-entrancy property, proving that a contract labeled secure by \ethor{} satisfies single-entrancy. This guarantee stems from the soundness of the abstract semantics with respect to the rigorously tested small step semantics and from the proof that call unreachability (formulated in terms of the abstract semantics) soundly approximates single-entrancy. 
The soundness of the abstract semantics further enables the sound verification of arbitrary reachability properties that are expressed in terms of the abstract semantics. In particular this holds for functional contract properties phrased as pre- and postconditions: \ethor{} can prove that a contract starting in a state satisfying the precondition is never able to reach a state that does not satisfy the postcondition.

}

\subsection{Implementation Strategies}
\REPLACECfor{FM}{200724}{
Instead of implementing the computationally intensive kernel of the analysis from scratch, the analysis task is usually reduced to a known problem supported by performant and well-maintained solvers.}
{To arrive at a fast and stable analysis implementation, the analysis task is usually reduced to a known problem supported by performant and well-maintained solvers.}
\REPLACECfor{FM}{200724}{Not only will this}{This does not only} save implementation time and help\NEWCfor{FM}{200724}{s} performance, but it also adds an abstraction layer that facilitates reasoning. \REMOVECfor{FM}{200724}{Further, when using a common output format like \smtlib{}~\cite{smtlib}, the implementation is not tied to a specific solver what provides further modularity.}
For \horst{}\REMOVECfor{FM}{200724}{, and therefore for \ethor{},} we decided to use \zz{}, respectively  Constrained Horn Clauses over \smtlib{}'s linear integer arithmetic fragment, as translation target.
We chose \zz{} since it is a state-of-the-art solver for program analysis and the fragment suffices to formulate reachability properties.

\MESSAGECfor{manual}{}{I think it could make the presentation easier if we would not try to discuss/give an overview on all possible ways of how one could potentially build a solver, but to stick with the problem/how to solve it in \ethor{} structure (see high-level comments)}

\subsubsection{Architecture}
\begin{figure}[t]
\centering{
\includegraphics[width=0.92\textwidth]{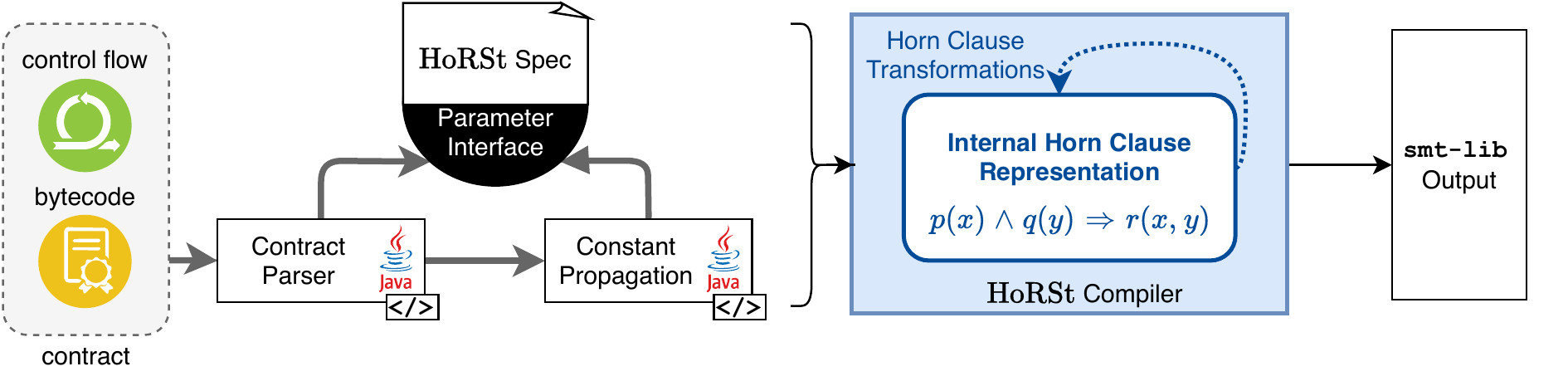}}
\caption{Architecture of \ethor{}}
\label{fig:ethor-architecture}
\end{figure}

In \ethor{}, the generation of \smtlib{} code that models the abstract semantics of a contract is structured into separate and well-defined phases.
As can be seen in \Cref{fig:ethor-architecture}, the input of \ethor{} consists of a contract with reconstructed control flow.
The bytecode of the contract is then parsed and constants are propagated within basic blocks.
With this information, the abstract semantics (provided as a \horst{} specification) is instantiated to a set of \hornc{} clauses which are, after several transformation steps, translated to \smtlib{} formulae.

\subsubsection{Optimizations}
\REPLACECfor{FM}{200724}{In order to be able to analyze real-world programs, it may be necessary to apply different optimizations.}{The performant analysis of real-world programs might require the usage of different optimizations, such as leveraging domain-specific knowledge in a pre-processing step.}
\REPLACECfor{FM}{200724}{
Given a concrete input, we can increase precision and performance by using domain knowledge in a preprocessing step.}
{}
Such preprocessing may include propagation of constants \cite{schneidewind2020ethor, tsankov_securify:_2018}, reconstruction of the control flow \cite{schneidewind2020ethor, tsankov_securify:_2018}, computation of memory offsets \cite{tsankov_securify:_2018}, and pruning of irrelevant parts of the input\cite{schneidewind2020ethor, tsankov_securify:_2018}. 
\MESSAGECfor{manual}{}{here you can simply back reference, since these works and their limitations are pretty much discussed now before}
\REMOVECfor{FM}{200724}{Depending on the used solver, it may also be helpful to apply transformations on the intermediate representation, such as  removing predicates \cite{schneidewind2020ethor}.}

As mentioned in \Cref{subsubsec:performance}, unsoundness introduced in any optimization or preprocessing step (e.g., by using an unsound decompiler) immediately affects the soundness of the whole analysis.
It is hence crucial to formally reason over each step.
\REPLACECfor{FM}{200724}{The control flow graph reconstruction of a smart contract in \ethor{}}{In \ethor{} the control flow graph reconstruction of a smart contract} is  realized by symbolically computing the destinations of all jump instructions based on a simplified version of the \NEWFfor{M}{200531}{sound} abstract semantics used for the later reachability analysis. Therefore, all soundness considerations from the full abstract semantics carry over to the preanalysis.
Since this version of the semantics falls into the datalog solvable fragment as implemented by the \souffle{} solver, we encoded this simple abstract semantics as a \souffle{} program. 
\REPLACECfor{FM}{200724}{In order to make it possible to generate preprocessing steps like this immediately from a declarative \horst{} specification that is suited for such reasoning, we are currently working to establish \souffle{} as an additional target for (a restricted subset of) \horst{}.}{
To automate the generation of such preprocessing steps in the future we plan to extend \horst{} with \souffle{} as additional compilation target.
}

\subsubsection{\REPLACECfor{FM}{200724}{Testing}{Evaluation}}
\NEWCfor{FM}{200724}{To ensure the correctness and perfomance of an analysis tool, it is inevitable to extensively and systematically test the tool implementation.}
\REPLACECfor{FM}{200724}{When developing a static analysis, s}{To this end s}ynthetic, well-understood inputs \REMOVECfor{FM}{200724}{can }help to identify problems regarding precision, performance, and correctness early. These, however, may not be representative of the challenges that are found in real-world contracts.
Data gathered from a real-world setting, on the other hand, might \REPLACECfor{FM}{200724}{be difficult to classify manually (i.e.\ check for presence or absence of properties), making it difficult to check for correctness of the implementation, and may overtax earlier, non-optimized iterations of the analysis tool.}{be of uncertain ground truth or too complex to give guidance in early stages of the development.}
In our experience, an automated test suite with corpus of synthetic and real-world inputs is a significant help while experimenting with different formulations and optimizations, as implementation bugs can be found already at an early stage.

For \ethor{} we leveraged the official EVM test suite and our own property-based test suite for assessing the correctness of \REPLACECfor{FM}{200724}{the semantics of single rules and simple contracts}{of the abstract semantics and abstract properties}. Out of $604$ relevant \NEWCfor{FM}{200724}{EVM} test cases, we terminated on 99\%\REPLACEFfor{CM}{200722}{, all of them confirming the tool's soundness.}{. All \REMOVECfor{FM}{200724}{of these} tests confirmed the tool's soundness and the possibility of specifying the test suite within \ethor{} confirmed the versatility of our approach beyond reentrancy.}
\REMOVECfor{FM}{200724}{We could show precise treatment (no overapproximations) in 83\% of the cases.}

\REPLACECfor{FM}{200724}{The assessment of correctness and precision for the \REPLACECfor{FM}{200724}{reentrancy}{single-entrancy} property \REMOVECfor{FM}{200724}{during development }was done with a}{The correctness and precision of \ethor{} for the single-entrancy property were assessed on a} benchmark of 712 real-world contracts. Within a 10 minute timeout, \ethor{} delivered results for 95\% of the contracts, with all of them confirming soundness, and yielding a specificity of 80\%, resulting in an F-meassure of 89\%. 
\NEWCfor{FM}{200724}{These results do not only demonstrate \ethor{}'s practicability on real-world contracts, but also clearly improve over the state-of-the-art analyzer ZEUS. When run on the same benchmark, ZEUS yields a specificity of only 11.4\% (challenging its soundness claim) and a specificity of 99.8\%, giving an F-measure of 20.4\%\footnote{\ethor{} was evaluated against ZEUS since this is the only tool to implement a property similar to single-entrancy.}}
\REMOVECfor{FM}{200724}{
\NEWFfor{CM}{200722}{Using the same benchmark for ZEUS, we assessed a specificity of only 11.4\% (challenging its soundness claim) and a specificity of 99.8\%, resulting in an F-measure of 20.4\%\NEWCfor{FM}{200724}{\footnote{\ethor{} was evaluated against ZEUS since this is the only tool to implement a property similar to single-entrancy.}} .} For the full experimental evaluation, we refer the reader to~\cite{schneidewind2020ethor}.}


\section{Future Challenges}

To bring forward the robust design and implementation of sound static analyzers, we plan on extending \horst{} in multiple ways: We want to integrate \horst{} with proof assistants in order to streamline and partially automate soundness proofs. Further, we want to add support for additional compilation targets, and enrich the specification language and compilation to go beyond reachability analysis, and to support restricted classes of hyperproperties.

For the particular case of \ethor{}, we want to improve the precision of the analysis, \REPLACECfor{FM}{200724}{in particular}{e.g.,} to include a symbolic treatment of hash values, and to enable the joint verification of multiple interacting contracts. Further, we strive to create a public benchmark of smart contracts exhibiting different security vulnerabilities, as well as mitigations. This would enable the community to systematically compare the performance, correctness, and precision of different tools.

Beyond that, we plan to transfer the presented techniques to other smart contract platforms, such a Libra, EOS, or Hyperledger Fabric, which exhibit domain-specific security properties and different semantics.

\section{Acknowledgements}

This work has been partially supported by the the European Research Council (ERC) under the European Union's Horizon 2020 research (grant agreement 771527-BROWSEC); by the Austrian Science Fund (FWF) through the projects PROFET (grant agreement P31621) and the project W1255-N23; by the Austrian Research Promotion Agency (FFG) through the Bridge-1 project PR4DLT (grant agreement 13808694) and the COMET K1 SBA; and by the Internet Foundation Austria (IPA) through the netidee project EtherTrust (Call 12, project 2158).

\bibliographystyle{splncs04}
\bibliography{ISoLA2020.bib}

\begin{thebibliography}{10}
\providecommand{\url}[1]{\texttt{#1}}
\providecommand{\urlprefix}{URL }
\providecommand{\doi}[1]{https://doi.org/#1}

\bibitem{thedao}
The {DAO} smart contract (2016), available at \url{http://etherscan.io/address/
  0xbb9bc244d798123fde783fcc1c72d3bb8c189413\#code}

\bibitem{paritya}
The parity wallet breach (2017), available at
  \url{https://www.coindesk.com/30-million-ether-reported-stolen-parity-wallet-breach/}

\bibitem{parityb}
The parity wallet vulnerability (2017), available at
  \url{https://paritytech.io/blog/security-alert.html}

\bibitem{solidity}
Solidity. https://solidity.readthedocs.io/ (2019)

\bibitem{adhikari2017secure}
Adhikari, C.: Secure framework for healthcare data management using
  ethereum-based blockchain technology  (2017)

\bibitem{amani2018towards}
Amani, S., B{\'e}gel, M., Bortin, M., Staples, M.: Towards verifying ethereum
  smart contract bytecode in isabelle/hol. In: Proceedings of the 7th ACM
  SIGPLAN International Conference on Certified Programs and Proofs. pp. 66--77
  (2018)

\bibitem{azaria2016medrec}
Azaria, A., Ekblaw, A., Vieira, T., Lippman, A.: Medrec: Using blockchain for
  medical data access and permission management. In: Open and Big Data (OBD),
  International Conference on. pp. 25--30. IEEE (2016)

\bibitem{bartoletti2019minimal}
Bartoletti, M., Galletta, L., Murgia, M.: A minimal core calculus for solidity
  contracts. In: Data Privacy Management, Cryptocurrencies and Blockchain
  Technology, pp. 233--243. Springer (2019)

\bibitem{bhargavan2016formal}
Bhargavan, K., Delignat-Lavaud, A., Fournet, C., Gollamudi, A., Gonthier, G.,
  Kobeissi, N., Kulatova, N., Rastogi, A., Sibut-Pinote, T., Swamy, N., et~al.:
  Formal verification of smart contracts: Short paper. In: Proceedings of the
  2016 ACM Workshop on Programming Languages and Analysis for Security. pp.
  91--96 (2016)

\bibitem{crafa2019solidity}
Crafa, S., Di~Pirro, M., Zucca, E.: Is solidity solid enough? In: International
  Conference on Financial Cryptography and Data Security. pp. 138--153.
  Springer (2019)

\bibitem{cruz2018rbac}
Cruz, J.P., Kaji, Y., Yanai, N.: Rbac-sc: Role-based access control using smart
  contract. Ieee Access  \textbf{6},  12240--12251 (2018)

\bibitem{de2008z3}
De~Moura, L., Bj{\o}rner, N.: Z3: An efficient smt solver. In: International
  conference on Tools and Algorithms for the Construction and Analysis of
  Systems. pp. 337--340. Springer (2008)

\bibitem{galal2018verifiable}
Galal, H.S., Youssef, A.M.: Verifiable sealed-bid auction on the ethereum
  blockchain. In: International Conference on Financial Cryptography and Data
  Security. pp. 265--278. Springer (2018)

\bibitem{GMS::CAV18}
Grishchenko, I., Maffei, M., Schneidewind, C.: Foundations and tools for the
  static analysis of ethereum smart contracts. In: Proceedings of the 30th
  International Conference on Computer-Aided Verification (CAV). pp. 51--78.
  Springer (2018)

\bibitem{GMS::POST18}
Grishchenko, I., Maffei, M., Schneidewind, C.: A semantic framework for the
  security analysis of ethereum smart contracts. In: Proceedings of the 7th
  International Conference on Principles of Security and Trust (POST). pp.
  243--269. Springer (2018)

\bibitem{grossman2017online}
Grossman, S., Abraham, I., Golan-Gueta, G., Michalevsky, Y., Rinetzky, N.,
  Sagiv, M., Zohar, Y.: Online detection of effectively callback free objects
  with applications to smart contracts. Proceedings of the ACM on Programming
  Languages  \textbf{2}(POPL),  1--28 (2017)

\bibitem{hahn2017smart}
Hahn, A., Singh, R., Liu, C.C., Chen, S.: Smart contract-based campus
  demonstration of decentralized transactive energy auctions. In: 2017 IEEE
  Power \& Energy Society Innovative Smart Grid Technologies Conference (ISGT).
  pp.~1--5. IEEE (2017)

\bibitem{hildenbrandt_kevm:_2018}
Hildenbrandt, E., Saxena, M., Rodrigues, N., Zhu, X., Daian, P., Guth, D.,
  Moore, B., Park, D., Zhang, Y., Stefanescu, A., Rosu, G.: {KEVM}: {A}
  {Complete} {Formal} {Semantics} of the {Ethereum} {Virtual} {Machine}. pp.
  204--217. IEEE (Jul 2018). \doi{10.1109/CSF.2018.00022},
  \url{https://ieeexplore.ieee.org/document/8429306/}

\bibitem{hirai2017defining}
Hirai, Y.: Defining the ethereum virtual machine for interactive theorem
  provers. In: International Conference on Financial Cryptography and Data
  Security. pp. 520--535. Springer (2017)

\bibitem{jiao2018executable}
Jiao, J., Kan, S., Lin, S.W., Sanan, D., Liu, Y., Sun, J.: Executable
  operational semantics of solidity. arXiv preprint arXiv:1804.01295  (2018)

\bibitem{jordan2016souffle}
Jordan, H., Scholz, B., Suboti{\'c}, P.: Souffl{\'e}: On synthesis of program
  analyzers. In: International Conference on Computer Aided Verification. pp.
  422--430. Springer (2016)

\bibitem{kalra_zeus:_2018}
Kalra, S., Goel, S., Dhawan, M., Sharma, S.: {ZEUS}: {Analyzing} {Safety} of
  {Smart} {Contracts}. Internet Society (2018). \doi{10.14722/ndss.2018.23082},
  \url{https://www.ndss-symposium.org/wp-content/uploads/2018/02/ndss2018{\_}09-1{\_}Kalra{\_}paper.pdf}

\bibitem{lu2019neucheck}
Lu, N., Wang, B., Zhang, Y., Shi, W., Esposito, C.: Neucheck: A more practical
  ethereum smart contract security analysis tool. Software: Practice and
  Experience  (2019)

\bibitem{luu2016making}
Luu, L., Chu, D.H., Olickel, H., Saxena, P., Hobor, A.: Making smart contracts
  smarter. In: Proceedings of the 2016 ACM SIGSAC conference on computer and
  communications security. pp. 254--269 (2016)

\bibitem{mathieu2017blocktix}
Mathieu, F., Mathee, R.: Blocktix: Decentralized event hosting and ticket
  distribution network  (2017), available at
  \url{https://blocktix.io/public/doc/blocktix-wp-draft.pdf}

\bibitem{mccorry2017smart}
McCorry, P., Shahandashti, S.F., Hao, F.: A smart contract for boardroom voting
  with maximum voter privacy. In: International Conference on Financial
  Cryptography and Data Security. pp. 357--375. Springer (2017)

\bibitem{nakamoto2008bitcoin}
Nakamoto, S.: Bitcoin: A peer-to-peer electronic cash system (2008), available
  at \url{http://bitcoin.org/bitcoin.pdf}

\bibitem{notheisen2017trading}
Notheisen, B., G{\"o}dde, M., Weinhardt, C.: Trading stocks on
  blocks-engineering decentralized markets. In: International Conference on
  Design Science Research in Information Systems. pp. 474--478. Springer (2017)

\bibitem{panescu2018smart}
Panescu, A.T., Manta, V.: Smart contracts for research data rights management
  over the ethereum blockchain network. Science \& Technology Libraries
  \textbf{37}(3),  235--245 (2018)

\bibitem{schneidewind2020ethor}
Schneidewind, C., Grishchenko, I., Scherer, M., Maffei, M.: ethor: Practical
  and provably sound static analysis of ethereum smart contracts. arXiv
  preprint arXiv:2005.06227  (2020)

\bibitem{tsankov_securify:_2018}
Tsankov, P., Dan, A., Drachsler-Cohen, D., Gervais, A., Bünzli, F., Vechev,
  M.: Securify: {Practical} {Security} {Analysis} of {Smart} {Contracts}. pp.
  67--82. ACM (Jan 2018). \doi{10.1145/3243734.3243780},
  \url{http://dl.acm.org/doi/10.1145/3243734.3243780}

\bibitem{wood2014ethereum}
Wood, G.: Ethereum: A secure decentralised generalised transaction ledger.
  Ethereum Project Yellow Paper  \textbf{151},  1--32 (2014)

\bibitem{wood_ethereum:2014}
Wood, G.: Ethereum: {A} secure decentralised generalised transaction ledger
  (2014)

\bibitem{yang2018lolisa}
Yang, Z., Lei, H.: Lolisa: Formal syntax and semantics for a subset of the
  solidity programming language. arXiv preprint arXiv:1803.09885  (2018)

\bibitem{zakrzewski2018towards}
Zakrzewski, J.: Towards verification of ethereum smart contracts: a
  formalization of core of solidity. In: Working Conference on Verified
  Software: Theories, Tools, and Experiments. pp. 229--247. Springer (2018)

\end{thebibliography}

\end{document}